\documentclass[12pt]{article}
\usepackage{graphicx}
\usepackage{subcaption}
\usepackage{array,dcolumn}
\usepackage{calc,tabularx, epsfig,mathrsfs}
\usepackage{hyperref}

\usepackage{amsmath,verbatim,enumerate}
\usepackage{amssymb}
\usepackage{multirow}
\usepackage{physics}
\usepackage[toc,page]{appendix}
\usepackage{xspace}
\usepackage{mathtools}
\usepackage{slashed}
\usepackage{cite}
\allowdisplaybreaks[1]

\usepackage{epstopdf}

\usepackage{setspace}
\newcommand{\beq}{\begin{equation}}
\newcommand{\eeq}{\end{equation}}
\newcommand{\bea}{\begin{eqnarray}}
\newcommand{\eea}{\end{eqnarray}}
\newcommand{\ba}{\begin{array}}
\newcommand{\ea}{\end{array}}
\newcommand{\bg}{\bar{g}}
\newcommand{\mn}{{\mu\nu}}

\newcommand{\pt}{\partial}

\newcommand{\hsf}{\hspace{5mm}}
\newcommand{\lt}{\left}
\newcommand{\rt}{\right}
\newcommand{\al}{\alpha}

\newcommand{\ep}{\epsilon}
\newcommand{\ta}{\theta}
\newcommand{\lam}{\lambda}
\newcommand{\Lam}{\Lambda}

\newcommand{\de}{\delta}
\newcommand{\D}{\Delta}
\newcommand{\OM}{\Omega}

\newcommand{\sg}{\sigma}

\begin{document}
\renewcommand{\theequation}{\thesection.\arabic{equation}}

\begin{titlepage}

  \bigskip\bigskip\bigskip\bigskip\bigskip

  \bigskip

  \vspace*{100px}

\centerline{\Large \bf {Lorentzian quantum cosmology with $R^2$ correction}}

    \bigskip

  \begin{center}

 \bf {Gaurav Narain$\,{}^1$ and Hai-Qing Zhang$\,{}^{1,2}$}
  \bigskip \rm
\bigskip

{\it  $1.$ Center for Gravitational Physics, Department of Space Science, 
Beihang University, Beijing 100191, China. \\
$2.$ International Research Institute for Multidisciplinary Science, Beihang University, Beijing 100191, China}
\smallskip

\vspace{1cm}
  \end{center}

  \bigskip\bigskip

 \bigskip\bigskip

%
\begin{abstract}
Quantum mechanical transition amplitudes directly tells the probability of each 
transition and which one is more favourable. Path-integrals 
offers a systematic methodology to compute this quantum mechanical process 
in a consistent manner. Although it is not complicated in simple quantum mechanical system 
but defining path-integral legitimately becomes highly nontrivial in the context of quantum-gravity,
where apart from usual issues of renormalizability, regularisation, measure, gauge-fixing,
boundary conditions, one still has to define the sensible integration contour for convergence. 
Picard-Lefschetz (PL) theory offers a unique way to find a contour of integration 
based on the analysis of saddle points and the steepest descent/ascent flow lines
in the complex plane. In this paper we make use of PL-theory to investigate 
Lorentzian quantum cosmology where the gravity gets modified in the ultraviolet with the 
$R^2$ corrections. We approach the problem perturbatively and compute the 
transition amplitude in the saddle point approximation to first order in higher-derivative 
coupling. This perturbative approximation is valid in certain regimes but the approximation 
cannot be used to address issues of very early Universe or no-boundary proposal. 
 \medskip
  \noindent
\end{abstract}

  \end{titlepage}

  \tableofcontents
%
%

\section{Introduction}
\label{intro}

Higher-derivative gravity theories have by now attained a respectable status in the area of 
modified theories of gravity (in context of inflation \cite{Starobinsky1980,Starobinsky1982,Maeda1987}) 
and as a well-defined renormalizable quantum field theory of gravity which is devoid of 
ultraviolet catastrophe \cite{Stelle:1976gc}. The existence of a nontrivial 
fixed point further adds to the merits of this simple modification of gravity on four spacetime 
dimensions \cite{Codello:2006in,Niedermaier:2009zz,Benedetti:2009rx,Groh:2011vn,Ohta:2015zwa}. 
The beauty offered here however comes at a cost of putting unitarity at stake
\cite{Salam:1978fd,Julve:1978xn}. 
Unitarity is an important ingredient in construction of successful 
field theories where the $S$-matrix is expected to follow optical theorem. 
Breakdown of unitarity is a hint that either vacuum decays (or vacuum instability), 
presence of tachyonic ghosts, or closure of Fock space. 

In some of recent works on investigation of higher-derivative gravity, effort was made 
to see whether quantum corrections can make the ghost unexcitable thereby making
it unphysical \cite{Salam:1978fd,Julve:1978xn}. It was indeed 
found that at least to 1-loop, the radiative corrections dictate the behaviour of flow of 
parameters in such a way so that the massive spin-2 ghost remains out of the physical 
spectrum irrespective of energy \cite{Narain:2011gs,Narain:2012nf,Narain:2013eea,Narain:2017tvp}
(same thing happens in scale-invariant higher-derivative gravity \cite{Narain:2016sgk}).
This led us to explore the impact of this result elsewhere:
energy regimes (cosmic era, spacetime regimes) where gravity might be getting modified.
Non-perturbative studies in Euclidean signature using functional renormalization group 
hints at the existence of a non-trivial fixed point in the theory space of couplings
thereby suggesting a possible UV completion. Whether this 
will hold in Lorentzian signature is yet to be established. A possibility exists as to do 
a Wick rotation sensibly and obtaining the Lorentizian case from Euclidean 
by properly implementing Wick rotation in curved spacetime 
\cite{Candelas:1977tt,Visser:2017atf,Baldazzi:2019kim,Baldazzi:2018mtl}. 
However, this direction is still in its infant stages and more work needs to be 
done.

The $R^2$ gravity, famously referred to as Starobinsky model
of inflation \cite{Starobinsky1980,Starobinsky1982}, is seen to posses 
wonderful properties as desired to have an early phase of cosmic inflation
and falls in the class of $F(R)$-theories of gravity 
\cite{Barrow1988,DeFelice2010}. 
Moreover, scalar field inflationary models via reconstruction 
can be cast into $F(R)$ theories and at high-energies are 
seen to become $R^2$ gravity \cite{Narain:2017mtu}. 
Motivated by acknowledging the supremacy of $R^2$ gravity among various 
inflationary models and the benefit of renormalizabilty it enjoys, we 
realise the necessity to investigate the effect of higher-derivative
gravity in the context of Lorentzian quantum cosmology. 

Flat spacetime has a meaningful time co-ordinate and enjoys the 
properties of global symmetries to cast Lorentz group in to 
a compact rotation group under a transformation of time co-ordinate. 
This is hard to replicate in a generic curved spacetime
which don't have such a symmetry group. 
In that sense Wick rotation (a process of defining a convergent 
path-integral by transforming the highly oscillatory path-integral in Lorentzian 
signature to euclidean) in quantum field theory on flat 
spacetime is more natural to implement than in curved spacetime
where `time' is just a parameter. The Feynman $+i\ep$-prescription 
in flat spacetime QFT offers a systematic way to make a choice of 
contour in complexified spacetime so that under Wick rotation 
the contour doesn't cross the poles of the free theory propagator. 
It not only offers necessary convergence in the path-integral but also 
implements causality in a systematic manner by requiring the 
euclideanised version of two-point function to satisfy Osterwalder-Schrader positivity. 
Generic lorentizian spacetime doesn't seem to offer any such restrictions
under analytic continuation of time co-ordinate. Things 
gets even harder to work out in cases when gravity is involved
{\it i.e} when background becomes dynamical or  
when gravity is treated as a quantum field. 

Application of Picard-Lefschetz theory has emerged as a promising approach 
to handle highly oscillatory integrals by integrating them along the 
path of steepest descent in the complexified plane 
which is uniquely figured out by making use of generalised flow 
equation in complex plane. Such steepest descent flow lines are 
termed \textit{Lefschetz thimbles}. Early attempts making use of  
knowledge of steepest descent contours occurred in the context of 
Euclidean quantum gravity \cite{Hawking:1981gb,Hartle:1983ai}
\beq
\label{eq:EucQGAct}
\int_{{\cal C}} {\cal D} g_\mn \exp \left(-I[g_\mn] \right) \, .
\eeq
The motivation to consider euclideanised version of path-integral for 
gravity was an expectation that similar to euclidean matter field theory on 
flat spacetime the gravitational theory will have the relevant convergence. 
Gravitational path-integral is a complicated business. Apart from the usual 
issues of path-integral measure, gauge-invariance (gauge-fixing),
regularization, renormalizability and boundary conditions; it is important 
to choose a contour of integration carefully for necessary convergence. 
This bit is in no way a trivial thing in curved spacetime and 
in particular when gravity is quantised. In flat spacetime the 
Feynman $+i\ep$-prescription (and the standard Wick rotation) save us 
from the complications; which no longer offers refuge in non-flat spacetime 
or in quantum gravity. 
 
Picard-Lefschetz theory offers a unique way of 
finding this integration contour in a generic spacetime where the 
gravitational path-integral becomes absolutely convergent. 
This has been noticed in the simple models of quantum cosmology 
\cite{Feldbrugge:2017kzv,Feldbrugge:2017fcc,Feldbrugge:2017mbc},
where the authors studied path-integral of Einstein-Hilbert gravity 
in the mini-superspace approximation. Earlier attempts in this direction 
(in Euclidean quantum cosmology)
goes back to 1980s \cite{Vilenkin:1982de,Vilenkin:1983xq,Vilenkin:1984wp,Hawking:1983hj}
when effort was being made to understand initial conditions for the cosmic 
inflation. It leads to \textit{tunnelling} proposal \cite{Vilenkin:1982de,Vilenkin:1983xq,Vilenkin:1984wp}
and \textit{no-boundary} proposal \cite{Hawking:1981gb,Hartle:1983ai,Hawking:1983hj}.
Here to make sense of euclidean path-integral of gravity (which is 
unbounded from below \cite{Gibbons:1977zz} due to famous conformal factor problem 
\cite{Gibbons:1978ac}) it is not enough to make a sensible initial 
condition choice but it was also noticed that additional input is needed 
to make a choice of contour of integration \cite{Halliwell:1988ik,Halliwell:1989dy,Halliwell:1990qr}.
Picard-Lefschetz theory allows one to pick the contour uniquely 
directly in Lorentzian spacetime and allows one to study
scenarios involving various initial conditions in a systematic manner
\cite{Feldbrugge:2017kzv,Feldbrugge:2017fcc,Feldbrugge:2017mbc}. 

In this paper we build on the foundation of Picard-Lefschetz theory 
to study the path-integral of gravity where gravity has been modified 
in the ultraviolet making it perturbatively renormalizable to all loops. 
Our aim is to consider eq. (\ref{eq:EucQGAct}) when requirements 
like gauge-fixing, regularization, renormalization, functional measure 
has been taken care of. This motivates us to tackle the case of 
higher-derivative gravity and consider the path-integral of the same 
in the mini-superspace approximation. We ask a straightforward 
question: what is the transition probability from one state to another, 
where the states are specified by the boundary conditions?
We address this in the context of mini-superspace approximation. 
\beq
\label{eq:minsup_ansat}
{\rm d}s^2 = - \frac{N^2(t)}{q(t)} + q(t) {\rm d}\OM_3^2
\eeq
where $N$ is lapse function and ${\rm d}\OM_3$ 
is the unit $3$-sphere metric. Our motivation to investigate the 
mini-superspace path-integral in the case of higher-derivative gravity is many folds:
(1) considering a perturbatively renormalizable 
gravity theory which in four spacetime is known to be
fourth order higher-derivative gravity \cite{Stelle:1976gc}
(2) Irrespective of whatever UV completion gravity may have, in effective 
field theory picture, gravity is expected to get modified, and the next leading 
modification is $R^2$ type. (3) Inflationary studies strongly advocates the 
$R^2$ model of inflation (4) The corresponding euclidean action is bounded
as $\exp(-\int {\rm d}x \,  \sqrt{g} R^2) \to 0$ as $\lvert R \rvert \to \infty$,
so fluctuations of metric resulting in large $\lvert R \rvert$ are suppressed. 

The outline of paper is following: in section \ref{hdgms} we discuss about the 
higher-derivative gravity mini-superspace model. In section \ref{eqmsoln}
we compute the solutions to equation of motion. In section \ref{TranProb}
we compute the transition probability. In section \ref{Ninnt} we perform the 
$N$-integration by making use of Picard-Lefschetz theory. 
We write the conclusion with outlook and discussion in 
section \ref{conc}.

\section{Higher-derivative mini-superspace}
\label{hdgms}

In this section we consider the higher-derivative gravity model 
which we take as Starobinsky model \cite{Starobinsky1980,Starobinsky1982}. 
The action for this is taken as
\beq
\label{eq:R2grav}
S_{\rm grav} = \frac{1}{16\pi G} \int {\rm d}^4x \sqrt{-g} \left(-2 \Lam + R + \al R^2 \right) \, ,
\eeq
where $R$ is the Ricci scalar of the spacetime metric $g_\mn$, $G$ 
is the Newton's gravitational constant (which has been pulled out),
$\Lam$ is the cosmological constant, 
while $\al$ is parameter dictating the strength of the higher-derivative term. 
$G$ has $[M]^{-2}$, $\Lam$ has mass dimension $[M]^{2}$
while $\al$ has mass dimension $[M]^{-2}$.
It is a Starobinsky model of gravity which has been well studied in the 
context of inflation and cosmology. From the effective field theory point of view 
the above action contains terms up to four derivatives of metric. 
In principle one should have an infinite tower of terms, however in 
four dimensions renormalizability is achieved with 
addition of $R^2$ and $R_\mn R^\mn$. 
In this paper we just study the Starobinsky model in the 
mini-superspace approximation and compute the 
probability of transition from one $3$-geometry to another in the 
path-integral formalism using Picard-Lefschetz (PL) theory. 
Although this seems like a straightforward extension of the 
work done previously in the context of Einstein-Hilbert gravity,
here addition of $R^2$ term is motivated due to renormalizabilty 
and its important role during early phase of Universe giving rise 
to deSitter expansion. 

The quantity that one is interested in computing is the transition 
probability from one geometry to another. It is a generalization of 
probability computation of usual quantum mechanics (or field theory)
to the case of gravity. In mini-superspace approximation this means
\beq
\label{eq:Gamp}
G[a,b]
= \int {\cal D} N(t)  \int_a^b {\cal D} q(t) \exp[i S_{\rm grav}/\hbar] \, , 
\eeq
where 
\beq
\label{eq:BC}
q(t=0)=a \, , 
\hspace{5mm} q(t=1)=b \, .
\eeq
The Ricci scalar $R$ for the metric given in 
eq. (\ref{eq:minsup_ansat}) is 
\beq
\label{eq:Rminsup}
R = \frac{3}{2N^3 q} \left(
4 k N^3 - 2 \dot{N} q \dot{q} + N \dot{q}^2 + 2 N q \ddot{q}
\right) \, ,
\eeq
where `dot' denotes time derivative, while $k$ is the curvature of the 
three dimensional space. The computation of the path-integral 
is a complicated task even in the mini-superspace approximation. The
usual complication regarding definition of measure, convergence,
un-controllable oscillations still remain in this approximation. 
Mostly in quantum mechanics the definition of measure is standard 
(by discretising), however often one obtains convergence following 
Feynman $+i\ep$-prescription or go to Euclidean via Wick rotation. 
The essence of Picard-Lefshetz is to offer a unique generalization 
of Feynman $+i\ep$-prescription and in the process it defines an 
absolutely convergent path-integral along the paths of steepest descent (called 
thimbles). WKB or saddle point approximation is a good 
strategy to compute a good approximation to the 
path-integral, where the saddle points are the solutions satisfying 
equation of motion. This is a boon in many cases as a particular 
classical system can be transformed in to another simpler classical system 
via canonical transformation where the system is easy to analyse. 
The two systems are equivalent on-shell however quantum 
mechanically they could be differ. 

The higher-derivative action that we consider falls in the category of 
$F(R)$-theories of gravity which are known to be classically equivalent to 
Einstein-Hilbert gravity coupled with scalar field \cite{Maeda1987,Barrow1988}. 
The two theories are related by a conformal transformation and are equivalent on-shell
\cite{DeFelice2010}. The action in Einstein-frame is given by
\beq
\label{eq:EFact}
S_{\rm E} = \int {\rm d} x \sqrt{-\bg}
\lt[
\frac{\bar{R}}{16 \pi G_E} - \frac{1}{2} \lt(\pt \phi \rt)^2 - V(\phi)
\rt] \, ,
\eeq
where $G_E$ is the Newton's constant in Einstein frame
and the metric in Einstein frame is related to metric in 
Jordon frame as $g_\mn = \OM^2 \bg_\mn$. 
If we set $(8\pi G_E)=1$ then we have 
\bea
\label{eq:pot}
&&
\OM^2 = e^{-\sqrt{2/3} \phi} \, ,
\notag \\
&&
V(\phi) = \left(\Lam + \frac{1}{4\al} \right) e^{-2\sqrt{2/3} \phi}
- \frac{1}{2\al} e^{-\sqrt{2/3} \phi} + \frac{1}{4\al} \, .
\eea
The saddle points which follow from equation of motion 
will be the same for the two theories described by action in 
eq. (\ref{eq:R2grav}) and (\ref{eq:EFact}). This knowledge 
can be exploited to compute the saddles of the theory using the 
simpler theory given in eq. (\ref{eq:EFact}) and use it to 
find saddles of theory in eq. (\ref{eq:R2grav}). For quantum 
theories it is hard to make a rigorous statement regarding the 
equivalence of the two. For now we assume that they may possibly differ.

In the mini-superspace approximation the metric in the Einstein-frame 
can be written as 
\beq
\label{eq:ef_minisup}
{\rm d}s^2 = - \frac{N_E^2(t)}{q_E(t)} + q_E(t) {\rm d}\OM_3^2 \, ,
\eeq
where the Lapse $N_E$ and $q_E$ are related to lapse $N$ 
and $q$ as follows (here the subscript $E$ denotes that corresponding 
quantities are in Einstein frame),
\beq
\label{eq:lapTr}
q_E(t) = e^{\sqrt{2/3} \phi} q(t) \, ,
\hsf
N_E = e^{\sqrt{2/3} \phi} N \, .
\eeq
In the gauge $\dot{N}_E=0$ one obtains the following action of the 
theory 
\bea
\label{eq:efminiscup_act}
&&
S_{\rm E} = V_3 \int {\rm d} t \biggl[
3 k N_E - \frac{3 \dot{q}_E^2}{4 N_E} 
+ \frac{q_E^2 \dot{\phi}^2}{2 N_E} 
\notag \\
&&
- N_E q_E \left\{
\left(\Lam + \frac{1}{4\al} \right) e^{-2\sqrt{\frac{2}{3}} \phi}
- \frac{1}{2\al} e^{-\sqrt{\frac{2}{3}} \phi} + \frac{1}{4\al}
\right\}
\biggr] \, ,
\eea
where the isotropicity of metric the field implies $\phi(x)=\phi(t)$.
This transformation work well enough as long as the $R^2$ coupling is 
non-zero. In the limit of $\al\to0$ the Jordon-frame theory reduces to 
known Einstein-Hilbert gravity, the conformal transformation 
and the potential becomes a constant. In the Jordon-frame the limit is more 
straightforward to take. In the following we will study the theory in the 
small $\al$ limit, considering the $R^2$ correction as a perturbation 
over the Einstein-Hilbert gravity.

\section{Jordon Frame solution}
\label{eqmsoln}

In Jordon frame one can also express the action of theory in mini-superspace 
approximation and proceed to find solution of equation of motion directly in 
Jordon frame. In principle one can conformally transform the theory 
in Jordon frame to Einstein frame and analyse the equation of motion 
there as the two theories are the same on-shell, however in doing that way 
it is not straightforward to cleanly take the $\al\to0$ limit. This motivate us 
analyse the solutions of theory in Jordon frame. 
The gravity action under the min-superspace 
approximation and in gauge $\dot{N}=0$ becomes following 
\bea
\label{eq:GravActMinisup}
&&
S_{\rm grav} = \frac{V_3}{2} \int {\rm d}t 
\biggl[
-2 \Lam N q + \frac{3}{2} \left\{
4 k N+ \frac{\dot{q}^2}{N} + \frac{2 q \ddot{q}}{N}
\right\}
\notag \\
&&
+ 9\al \left\{
\frac{4 k^2 N}{q} + \frac{\dot{q}^4}{4N^3 q} 
+ \frac{q \ddot{q}^2}{N^3} + \frac{2 k \dot{q}^2}{Nq} 
+ \frac{\dot{q}^2 \ddot{q} }{N^3} + \frac{4 k \ddot{q}}{N}
\right\}
\biggr] \, ,
\eea
where we have absorbed $8 \pi G$ in the $3$-dimensional volume of space. 
Note the presence of terms which goes like $\sim 1/q$ in the contribution
coming from $R^2$ piece. This is merely an indication that at small $q$
the higher-derivative gravity piece starts to play an important role. 
Ignoring surface terms one can obtain an equation of motion by doing a first variation 
of this action with respect to $q(t)$. Computation of this is an easy task but finding a
non-perturbative solution of the same is unknown. This equation is given by,
\bea
\label{eq:eqmact}
&&
- 2 \Lam N + \frac{3 \ddot{q}}{N} 
+ \frac{9 \al}{N^3} \biggl[
3 \ddot{q}^2 + 4 \dot{q} \dddot{q} + 2 q \ddddot{q}
+ \frac{2 k N^2 \dot{q}^2}{q^2} - \frac{4 k N^2\ddot{q}}{q} 
- \frac{3 N^2 \dot{q}^2 \ddot{q}}{q} 
\notag \\
&&
+ \frac{3 N^2 \dot{q}^4}{4 q^2} 
- \frac{4 k^2 N^4}{q^2}
\biggr] = 0 \, .
\eea
Here the second line contain terms coming from higher-derivatives. This piece 
not only has four time-derivatives of $q$ but also is non-linear in nature, thereby 
making the system highly complicated to solve. This equation one has to solve 
along with the boundary conditions imposed on $q(t)$ given in eq. (\ref{eq:BC}). 
The equation also has terms going like $\sim 1/q$ which will become significantly 
important in the small $q\to0$ limit, the case corresponding to no-boundary proposal.
Such terms don't appear in the pure EH gravity (or EH gravity coupled with scalar). 
Cases where its absence gave rise 
to a possibility of having a no-boundary Universe: a Universe starting from nothing. 
In $R^2$ modification of gravity this seems to be questionable due to presence 
of $1/q$ terms in equation of motion. Addressing this issue requires 
a non-perturbative study of the system, a perturbative analysis won't be 
able to tackle this issue. Perturbative analysis has a regime of validity,
beyond which (small $q$) it breaks down. 

$R^2$ modified gravity is complicated  
but it offers more richness in the gravitational system and may give rise to 
new saddles which were not present in EH gravity. By varying the action 
with respect to $N$ we notice that we get a constraint 
which basically imposes a condition of the solution $q(t)$. This constraint is given by,
\bea
\label{eq:constNq}
- 2 \Lam q + \frac{3}{2} \left(4 k + \frac{\dot{q}^2}{N^2} \right)
+ \frac{9 \al}{4 N^4} \left\{
-12 q \ddot{q}^2 - \frac{8 k N^2 \dot{q}^2}{q^2} - \frac{3 \dot{q}^4}{q^2} + \frac{16 N^4}{q^2}
\right\} = 0 \, .
\eea
%

\subsection{Constant solution}
\label{consSad}

We first try whether the equation of motion has a constant solution. We set 
$q(t)=cons.$, which means any $t$-derivative of $q(t)$ is zero. It is possible to 
have a constant solution if in the boundary condition $a=b$. 
The equation of motion simplifies to
\beq
\label{eq:eqmCons}
N\left(\Lam + \frac{18 \al k^2}{q^2} \right) =0 \, .
\eeq
For $N\neq0$ this has two complex conjugate solutions 
\beq
\label{eq:consSoln}
q(t) = \pm \sqrt{-\frac{18 \al k^2}{\Lam}} \, ,
\eeq
where to satisfy the boundary condition we have 
\beq
\label{eq:BCeqCo}
a=b=\pm \sqrt{-\frac{18 \al k^2}{\Lam}} \, ,
\eeq
where we know that as $a$ and $b$ are real and positive,
so this implies $\Lam<0$ and in the solution for $q(t)$
one takes the positive sign. This also fixes $\Lam = -18 \al k^2 /a^2$. 
The Ricci scalar 
\beq
\label{eq:Rconst}
R = \frac{6k}{a} \, .
\eeq
This is trivial solution to the system.

\subsection{Perturbative solution}
\label{perturbSol}

One can solve the eq. (\ref{eq:eqmact}) perturbatively in $\al$, where we assume 
that higher-derivative corrections if any exists must be small. In effective field theory 
approach this is a valid and reasonable assumption, where one can make computation 
of transition probabilities perturbatively. This works well as long as the perturbation 
theory doesn't break down and higher-derivative remains small in the regime of study. 
In this subsection we work in this regime and investigate the problem perturbatively. 
Later we comment on the situation when this no longer holds. 
One can write the solution in powers of $\al$ as
\beq
\label{eq:solnAL}
q(t) = q_0(t) + \al q_1(t) + {\cal O}(\al^2) \, .
\eeq
Plugging this in the above equation we notice that $q_0(t)$ and $q_1(t)$ satisfy the 
following system of equations
\begin{align}
\label{eq:q0}
&
- 2 \Lam N + \frac{3 \ddot{q_0}}{N} = 0 \, ,
\\
&
\label{eq:q1}
\ddot{q_1}
+ \frac{3}{N^2} \biggl[
3 \ddot{q_0}^2 + 4 \dot{q_0} \dddot{q_0} + 2 q \ddddot{q_0}
+ \frac{2 k N^2 \dot{q_0}^2}{q_0^2} - \frac{4 k N^2\ddot{q_0}}{q_0} 
- \frac{3 N^2 \dot{q_0}^2 \ddot{q_0}}{q_0} + \frac{3 N^2 \dot{q_0}^4}{4 q_0^2} 
- \frac{4 k^2 N^4}{q_0^2}
\biggr] = 0 \, .
\end{align}
These two equations can now be solved by making use of the boundary conditions
given by the requirement that at each order in $\al$ the boundary conditions 
given in eq. (\ref{eq:BC}) are to be satisfied. 
Solution of the former gives the following 
\beq
\label{eq:q0soln}
q_0(t) = \frac{\Lam N^2}{3} (t^2 -t) + b t + a 
= \frac{\Lam}{3} N^2 (t-p_+)(t-p_-)\, ,
\eeq
where we have factored the quadratic quadratic polynomial in $t$ 
whose roots are denoted as 
\beq
\label{eq:rootsppm}
p_\pm = \frac{1}{2} + \frac{3(a-b)}{2 N^2 \Lam}
\pm \frac{3}{2N^2 \Lam} \sqrt{
\left(a-b+\frac{N^2 \Lam}{3} \right)^2 - \frac{4 a N^2 \Lam}{3}
} \, .
\eeq
Notice here the importance of term $\Lam$, had it been absent in our action 
then the $q_0(t)$ is independent of $N$. Presence of $\Lam$ introduces 
$N^2$ dependences in the roots $p_\pm$.
Solving for $q_1$ is a straightforward but a lengthy algebra. The solution 
obtained for $q_0(t)$ can be plugged in the equation for $q_1(t)$. The boundary 
conditions for $q_1(t)$ is obtained by requiring that the full $q(t)$ should satisfy 
eq. (\ref{eq:BC}). This implies 
\beq
\label{eq:BCq1}
q_1(t=0) = q_1 (t=1) =0 \, . 
\eeq
$q_1(t)$ obeys a second order linear ODE with above boundary conditions. 
The system can be solved easily giving a lengthy expression for $q_1(t)$. 
After the lengthy algebra we obtain the following solution for $q_1(t)$
\bea
\label{eq:q1soln}
&&
q_1(t) = r_1 t + r_2 -32 \Lam^2 N^4 (p_+ + p_{-}) t - 16 \Lam^2 N^2(1+2N^2) t^2 
- \biggl[
32 \Lam^2 N^4 (p_+-p_{-}) 
\notag \\
&&
- \frac{12 k}{(p_+-p_{-})} + \frac{108 k}{\Lam^2 N^2 (p_+-p_{-})^3}
\biggr] \left\{(t-p_+) \log(p_+-t) - (t-p_-) \log(p_--t)\right\}
\notag \\
&&
- \biggl[
\frac{54k}{\Lam^2 N^2 (p_+ - p_-)^2} - 8 \Lam^2 N^4 (p_+ - p_-)^2
\biggr] \log \left\{(t-p_+)(t-p_-) \right\} 
\notag \\
&&
+12 k \log q_0(t) \, .
\eea
where $r_{1,2}$ are integrations constants which are determined using the 
boundary conditions stated in eq. (\ref{eq:BCq1}). The expressions for these are 
quite lengthy. The important thing to note in the solution of $q_1$ is that it 
also contain terms like $\log N$.
\bea
\label{eq:q1cons}
&&
r_1 = 16 \Lam^2 N^2 (1+2N^2) + 32 \Lam^2 N^4 (p_++p_-) + 12 k \log (a/b)
\biggl[
\frac{24k}{p_+ - p_-} 
\notag \\
&&
- 32 \Lam^2 N^4 (p_+ - p_-) 
- \frac{108 k}{\Lam^2 N^2 (p_+ - p_-)^3}
\biggr] \biggl\{
(1-p_+) \log(p_+-1) + p_+ \log p_+ 
\notag \\
&&
- (1-p_-) \log(p_--1) - p_- \log p_-
\biggr\}
\notag \\
&&
+ \biggl[
\frac{54k}{\Lam^2 N^2 (p_+ - p_-)^2} - 8 \Lam^2 N^4 (p_+ - p_-)^2
\biggr] \log \frac{(1-p_+)(1-p_-)}{p_+ p_-} \, ,
\notag \\
&&
r_2 = \biggl[
\frac{24k}{p_+ - p_-} - 32 \Lam^2 N^4 (p_+ - p_-) - \frac{108 k}{\Lam^2 N^2 (p_+ - p_-)^3}
\biggr] (p_+ \log p_+ - p_-\log p_-) 
\notag \\
&&
-12 k \log a
+ \biggl[
\frac{54k}{\Lam^2 N^2 (p_+ - p_-)^2} - 8 \Lam^2 N^4 (p_+ - p_-)^2
\biggr] \log \left(\frac{3a}{2\Lam N^2}\right) \, .
\eea
From this it follows that the solution contains terms like $\log N$ 
and $\log a$.This could be artefact of the perturbative nature of study,
where emergence of $\log a$ follows from solving the equation of motion
perturbatively in $\al$. A direct implication of this is that the limit $a\to0$ is 
questionable. But then such inferences can't be reliably made 
in perturbative treatment. These kind of terms were 
definitely not present in the EH case studied in \cite{Feldbrugge:2017kzv,Feldbrugge:2017fcc,Feldbrugge:2017mbc}.
In EH case the simplicity of system allowed a non-perturbative analysis.

The full solution of $q(t)$ will satisfy both the equation of motion and the above constraint. 
This will imply that in general we have 
\beq
\label{eq:qdecomp}
q(t) = q_b(t) + Q(t) \, ,
\eeq
where $q_b(t)$ satisfies the equation of motion while 
$Q(t)$ is the fluctuation around the background $q_b$. The on-shell action 
to order ${\cal O}(\al)$ is given by 
\bea
\label{eq:S0Oal}
&&
S^{(0)}_{\rm grav} = \frac{V_3}{2} \int_0^1 {\rm d}t \biggl[
-2 \Lam N (q_0 + \al q_1)  
+ \frac{3}{2} \left\{
4 k N - \frac{\dot{q_0}^2}{N} - 2 \al \frac{\dot{q_0} \dot{q_1}}{N} 
\right\}
\notag \\
&&
+ \frac{9 \al}{N^3} \left\{
q_0 \ddot{q_0}^2 + 2 k N^2 \frac{\dot{q_0}^2}{q_0} 
+ \frac{\dot{q_0}^4}{4 q_0^2} + \frac{4 k^2 N^4}{q_0}
\right\}
\biggr] \, .
\eea
If one plugs the expression for $q_0$ and $q_1$ and integrate the time co-ordinate, 
then we get a complicated expression for the on-shell action to first order in $\al$. 
The $t$-integration is involved as the expression for both $q_0$ and $q_1$
has a non-trivial structure. It is interesting to note that after plugging the 
solutions the following contribution 
\beq
\label{eq:noteCond}
\int_0^1 {\rm d} t \left(-2 \Lam N q_1 - 3 \dot{q}_0 \dot{q}_1/N \right) =0.
\eeq
This simplifies our algebra leaving behind an expression for 
on-shell action which is functional only of $q_0$ and its 
time derivatives. 
\beq
\label{eq:S0Oal1}
S^{(0)}_{\rm grav} = \frac{V_3}{2} \int_0^1 {\rm d}t \biggl[
6 k N
-2 \Lam N q_0   
- \frac{3\dot{q_0}^2}{2N} 
+ 9 \al \left(
\frac{q_0 \ddot{q_0}^2}{N^3} + \frac{2 k \dot{q_0}^2}{N q_0} 
+ \frac{\dot{q_0}^4}{4 N^3 q_0^2} + \frac{4 k^2 N}{q_0}
\right)
\biggr] \, .
\eeq
The action $S^{(0)}_{\rm grav}$ after $t$-integration is given by
\bea
\label{eq:onshellTint}
&&
S^{(0)}_{\rm grav} =  \frac{V_3}{2} \biggl[
-\frac{3(a-b)^2}{2N} + 6kN - (a+b) N\Lam + \frac{N^3 \Lam^2}{18} 
+ \frac{9\al}{4ab} \biggl\{
\frac{(a-b)^4}{N^3}
\notag \\
&&
-\frac{(a-b)^2(a+b)\Lam}{N}
+ \frac{N\left(96abk \Lam + (3a^2 + 3b^2 + 18 ab + 8ab^2 + 8a^2 b) \Lam^2 \right)}{9}
\notag \\
&&
- \frac{(3a+3b + 8ab) \Lam^3 N^3}{81} \biggr\}
+ \frac{3\al}{N \sqrt{-L(N)}} \biggl(
\tan^{-1} \frac{3(a-b) + N^2 \Lam}{\sqrt{-L(N)}}
\notag \\
&&
+ \tan^{-1} \frac{3(b-a) + N^2 \Lam}{\sqrt{-L(N)}}
\biggr)
\biggl\{
(4k +\Lam) L(N) + 72 k^2 N^2
\biggr\}
\biggr] \, ,
\eea
where
\beq
\label{eq:QuadN}
L(N) = 9(a-b)^2 - 6ab N^2 \Lam + N^4 \Lam^2 \, ,
\eeq
In the following we will make use of this to compute the transition probability 
in the saddle point approximation. 

\section{Transition Probability}
\label{TranProb}

The main aim of the work is to obtain an expression for the transition probability from 
one geometry to another, which we do it perturbatively in the higher-derivative coupling.
The process of obtaining this leads to an evaluation of path-integral in mini-superspace
approximation which is divided in two parts: an integral over $N$ 
(in the constant gauge $\dot{N}=0$) and path-integral over $q(t)$ with the 
boundary conditions satisfied. The later can be evaluated along the lines 
of WKB where the leading contribution comes from the saddle points
with corrections coming from fluctuations. Expressing $q(t)$ as 
in eq. (\ref{eq:qdecomp}), the mini-superspace gravity action can be expanded in powers of $Q(t)$. 
This will generate an infinite tower of terms due to the highly non-linear nature of the action. 
The second variation of action is given by
\bea
\label{eq:2ndvarAct}
&&
S^{(2)}_{\rm grav} = \frac{V_3}{2} \int_0^1 {\rm d}t \biggl[
\left\{ -\frac{3}{2N} - \frac{18\al}{N^3} \left(\ddot{q_b} - \frac{k N^2}{q_b} 
- \frac{3 \dot{q_b}^2}{4 q_b^2} \right) \right\} \dot{Q}^2
+ \frac{9 \al q_b}{N^3} \ddot{Q}^2 
\notag \\
&&
+ \frac{9\al}{N^3} \left\{
q_b^{(4)} + 2 k N^2 \frac{\ddot{q_b}}{q_b^2} - 2k N^2 \frac{\dot{q_b}^2}{q_b^3} 
+ \frac{3 \dot{q_b}^2 \ddot{q_b}}{2 q_b^3} 
- \frac{5 \dot{q_b}^4}{4q_b^4} + \frac{16 k^2 N^4}{9q_b^3} 
\right\}Q^2 + \cdots
\biggr] \, .
\eea
Here $q_b$ is the background which obeys the equation of motion. 
In obtaining this expansion we haven't made an assumption regarding 
$\al$ to be small, as the $q_b(t)$ entering here is a full solution 
of the equation of the motion. In Jordon frame we solve for $q_b(t)$
perturbatively in $\al$ retaining terms to first order in $\al$. If one 
writes $q_b = q_0 + \al q_1$, then to first order in $\al$ the second variation 
mentioend in eq. (\ref{eq:2ndvarAct}) gets further simplified. 
The path-integral in eq. (\ref{eq:Gamp}) reduces to 
following where we keep terms to first order in $\al$. This is given by
\bea
\label{eq:S2firstOal}
&&
S^{(2)}_{\rm grav} = 
S^{(2)}_{\rm EH} + S^{(2)}_{\rm HDG}=
\frac{V_3}{2} \int_0^1 {\rm d}t \biggl[
\left\{ -\frac{3}{2N} - \frac{18\al}{N^3} \left(\ddot{q_0} - \frac{k N^2}{q_0} 
- \frac{3 \dot{q_0}^2}{4 q_0^2} \right) \right\} \dot{Q}^2
+ \frac{9 \al q_0}{N^3} \ddot{Q}^2 
\notag \\
&&
+ \frac{9\al}{N^3} \left\{
2 k N^2 \frac{\ddot{q_0}}{q_0^2} - 2k N^2 \frac{\dot{q_0}^2}{q_0^3} 
+ \frac{3 \dot{q_0}^2 \ddot{q_0}}{2 q_0^3} 
- \frac{5 \dot{q_0}^4}{4q_0^4} + \frac{16 k^2 N^4}{9q_0^3} 
\right\}Q^2 + {\cal O}(\al^2)
\biggr] \, ,
\eea
where $S^{2)}_{\rm HDG}$ refers to higher-derivative gravity part.
After the decomposition we get the following form of the path-integral. 
\beq
\label{eq:pathDecomp}
G[a,b]
= \int_{0^+}^{\infty} {\rm d} N \exp\left( \frac{iS^{(0)}_{\rm grav}}{\hbar}\right)
\int_{Q[0]=0}^{Q[1]=0} {\cal D} Q(t) \exp\left(\frac{i S^{(2)}_{\rm grav}}{\hbar}\right) \, . 
\eeq
Now the crucial part is performing the path-integration over fluctuation 
$Q(t)$ which vanishes at both boundaries. In the case of Einstein-Hilbert 
gravity the form of this path-integral is very simple and it is easy to evaluate 
it exactly. Moreover, in case of EH gravity the expansion in powers of $Q(t)$ stops 
at quadratic order, making the path-integral over $Q(t)$ to be nice gaussian-integral. 
In case of higher-derivative gravity this doesn't happen. Even when the 
power series is truncated at second order in $Q$, still it has a complicated structure. 
In this paper we evaluate the path-integral over $Q(t)$ perturbatively to first order in $\al$.

\subsection{$Q$-integration}
\label{Qint}

The computation of the $Q$-integration has to be done perturbatively largely because 
the exponential appearing in the integrand is complicated, and since we have 
obtained background solutions perturbatively in $\al$ therefore for consistency 
it is required that we perform the path-integral perturbatively. Still being 
quadratic in $Q$ (as we retain terms upto second order in $Q$), it may appear 
that it should be possible to do it without resorting to approximations
as the second variation has the structure 
\beq
\label{eq:S2struct}
S^{(2)} = \int {\rm d}t \left[
\left(B_0 + B_1(t) \right) \dot{Q}^2 + B_2(t) \ddot{Q}^2 + B_3(t) Q^2
\right] \, ,
\eeq
where $B_0$ is constant in $t$, while rest of the coefficients are 
time-dependent. This $t$-dependence results in complications. If these 
coefficients were constants then it should be possible to perform the 
path-integral over $Q$ exactly. This we cannot do now, and we therefore 
do the computation perturbatively in $\al$ to first order. 

We first note that the fluctuation $Q(t)$ vanishes at the two boundary points.
This mean that it has following decomposition
\beq
\label{eq:Qdecomp}
Q(t) = \sum_{\lvert k \rvert \geq 1}^{\infty} c_k \exp \left(2 \pi k t \right) \, ,
\hspace{5mm}
{\rm with}
\hspace{5mm}
c_{-k} = c^*_k \, .
\eeq
This implies that the path-integral measure accordingly becomes the 
following
\beq
\label{eq:meaDecomp}
{\cal D} Q(t) = {\cal N} \int_{-\infty}^{\infty} \prod_{\lvert k \rvert \geq 1}^{\infty} {\rm d} c_k \, ,
\eeq
where the normalisation ${\cal N}$ needs to be determined carefully. 
In path-integral one chooses the normalisation in such a manner 
so that the end result is finite and doesn't contain divergences 
coming from infinite summation or product over integer $k$. 
In case of Einstein-Hilbert gravity one has a simpler
path-integral for $Q$ to be performed which is given by
\beq
\label{eq:pathQEH}
\int_{Q[0]=0}^{Q[1]=0} {\cal D} Q(t) \exp\left(\frac{i S^{(2)}_{\rm EH}}{\hbar}\right)
= \int_{Q[0]=0}^{Q[1]=0} {\cal D} Q(t) \exp\left(
-\frac{3 i V_3}{4N \hbar} \int_{0}^1 {\rm d} t \dot{Q}^2
\right) \, .
\eeq
This is similar to a free particle path-integral whose evaluation is 
graduate textbook exercise. Its value is actually finite.
However, when one insert the decomposition for $Q(t)$ given in 
eq. (\ref{eq:Qdecomp}) and write the measure as in eq. (\ref{eq:meaDecomp}),
and perform the path-integral then one encounters infinities. 
\beq
\label{eq:meaEH}
{\cal N}_{\rm EH} \int_{-\infty}^{\infty} 
\prod_{\lvert k \rvert \geq 1}^{\infty} {\rm d} c_k
\exp \left(
\frac{3 i V_3}{4 N \hbar} \sum_k (2\pi k)^2 \lvert c_k \rvert^2
\right)
= \left(
\frac{3 i V_3}{4 \pi N \hbar}
\right)^{1/2} \, .
\eeq
The LHS is an infinite product of gaussian integrals. This will have 
infinities which can be absorbed by appropriately defining 
${\cal N}_{\rm EH}$. This will give basically 
\beq
\label{eq:Neh_exp}
{\cal N}_{\rm EH} \prod_{k=1}^{\infty} \frac{8 \pi N \hbar}{3 (2 \pi k)^2 V_3} 
= \left(
\frac{3 i V_3}{4 \pi N \hbar}
\right)^{1/2} \, .
\eeq
In the case of higher-derivative gravity the normalisation needs to be 
fixed accordingly. At this point we also write $c_n = a_n + i b_n$
and $c_{-n} = c_n^* = a_n - i b_n$, where $a_n$ and $b_n$ are real. 
This result of change of variable will introduce a jacobian factor. 
The gravity action consist of two parts: $S^{(2)}_{\rm grav}
= S^{(2)}_{\rm EH} + S^{(2)}_{\rm HDG}$. 
As were are working perturbatively in coupling $\al$, so this implies
\bea
\label{eq:hdgpath_exp}
&&
\int_{Q[0]=0}^{Q[1]=0} {\cal D} Q(t) \exp\left(\frac{i S^{(2)}_{\rm grav}}{\hbar}\right)
= {\cal N}_{\rm EH} \left(1 + \al {\cal N}_1 + \cdots \right)
\notag \\
&&
\times
\int_{-\infty}^{\infty} \prod_{k=1}^{\infty} {\rm d} a_k {\rm d} b_k \left(\frac{2}{i}\right)
\left(1 + \frac{i}{\hbar} S^{(2)}_{\rm HDG} + \cdots \right)
\exp\left(\frac{i S^{(2)}_{\rm EH}}{\hbar}\right) \, ,
\notag \\
&&
= {\cal N}_{\rm EH} \int_{-\infty}^{\infty} \prod_{k=1}^{\infty} {\rm d} a_k {\rm d} b_k \left(\frac{2}{i}\right)
\exp\left(\frac{i S^{(2)}_{\rm EH}}{\hbar}\right)
\notag \\
&&
+ {\cal N}_{\rm EH} \int_{-\infty}^{\infty} \prod_{k=1}^{\infty} {\rm d} a_k {\rm d} b_k \left(\frac{2}{i}\right)
\biggl[
\al {\cal N}_1+ \frac{i}{\hbar} S^{(2)}_{\rm HDG} 
\biggr] \exp\left(\frac{i S^{(2)}_{\rm EH}}{\hbar}\right)
+ {\cal O}(\al^2) \, ,
\eea
where ${\cal N}_{\rm EH}$ is given in eq. (\ref{eq:Neh_exp}), 
${\cal N}_1$ is the infinite constant which will be adjusted to 
absorb the infinity coming from HDG, factor $2/i$ arises 
due to jacobian transformation. The EH action is 
quadratic in $a_n$ and $b_n$, after we plug the decomposition 
for $Q(t)$ and integrate with respect to time. 
$S^{(2)}_{\rm HDG}$ on the other hand will contain mixed terms 
too for example like $a_m a_n$, $a_m b_n$, and $b_m b_n$. 
Such kind of terms remain even after the $t$-integration. This is 
due to non-trivial dependence of second variation on $q_0(t)$ and its derivatives. 
The $S^{(2)}_{\rm HDG}$ is given by,
\bea
\label{eq:HDGS2_cns}
&&
S^{(2)}_{\rm HDG} = \frac{V_3}{2}  \sum_{\lvert k, k^\prime \rvert \geq 1}^{\infty}
\int_0^1 {\rm d}t 
\biggl[
\frac{18\al}{N^3}  \left(\ddot{q_0} - \frac{k N^2}{q_0} 
- \frac{3 \dot{q_0}^2}{4 q_0^2} \right) (4\pi^2 k k^\prime)
+ \frac{9 \al q_0}{N^3} (16 \pi^4 k^2 k^{\prime 2}) 
\notag \\
&&
+ \frac{9\al}{N^3} \left(
2 k N^2 \frac{\ddot{q_0}}{q_0^2} - 2k N^2 \frac{\dot{q_0}^2}{q_0^3} 
+ \frac{3 \dot{q_0}^2 \ddot{q_0}}{2 q_0^3} 
- \frac{5 \dot{q_0}^4}{4q_0^4} + \frac{16 k^2 N^4}{9q_0^3} 
\right)
\biggr] c_k c_{k^\prime} e^{2\pi i(k+k^\prime)t} \, ,
\eea
where $q_0(t)$ is quadratic in $t$ and is given in eq. (\ref{eq:q0soln}). Here 
one has to do $t$-integration. Writing $c_k$'s in terms of $a_k$'s and 
$b_k$'s, it is possible to write the above expression as a summation 
over only positive integer values of $k$ and $k^\prime$. The resulting 
expression will also contain mixed terms. 
Such kind of terms are non-diagonal. But as we are doing a
gaussian integral where the exponential is quadratic in
$a_k$ and $b_k$ (which is even function), it therefore 
implies that any kind of mixed term will not contribute. 
We introduce a shorthand 
\bea
\label{eq:Mkkshort}
&&
M(k,k^\prime) = 
\int_0^1 {\rm d}t 
\biggl[
\frac{18\al}{N^3}  \left(\ddot{q_0} - \frac{k N^2}{q_0} 
- \frac{3 \dot{q_0}^2}{4 q_0^2} \right) (4\pi^2 k k^\prime)
+ \frac{9 \al q_0}{N^3} (16 \pi^4 k^2 k^{\prime 2}) 
\notag \\
&&
+ \frac{9\al}{N^3} \left(
2 k N^2 \frac{\ddot{q_0}}{q_0^2} - 2k N^2 \frac{\dot{q_0}^2}{q_0^3} 
+ \frac{3 \dot{q_0}^2 \ddot{q_0}}{2 q_0^3} 
- \frac{5 \dot{q_0}^4}{4q_0^4} + \frac{16 k^2 N^4}{9q_0^3} 
\right)
\biggr] e^{2\pi i(k+k^\prime)t} \, ,
\notag\\
&&
= \int_0^1 {\rm d}t 
\left[
4\pi^2 k k^\prime A_1(t) + 16\pi^4 k^2 k^{\prime 2} A_2(t)  +  A_3(t) 
\right] e^{2\pi i(k+k^\prime)t} \, ,
\eea
where we have 
\begin{align}
\label{eq:A1}
&
A_1(t)= \frac{18\al}{N^3} \left(\ddot{q_0} - \frac{k N^2}{q_0} 
- \frac{3 \dot{q_0}^2}{4 q_0^2} \right) \, , \\
\label{eq:A2}
&
A_2(t) = \frac{9 \al q_0}{N^3}  \, , \\
\label{eq:A3}
&
A_3(t) = \frac{9\al}{N^3}  
\left(2 k N^2 \frac{\ddot{q_0}}{q_0^2} - 2k N^2 \frac{\dot{q_0}^2}{q_0^3} 
+ \frac{3 \dot{q_0}^2 \ddot{q_0}}{2 q_0^3} 
- \frac{5 \dot{q_0}^4}{4q_0^4} + \frac{16 k^2 N^4}{9q_0^3} 
\right)  \, .
\end{align}
This shorthand is useful as it expresses the structure of the 
$S^{(2)}_{\rm HDG}$ in a simple manner. This is given by,
\bea
\label{eq:S2hdgshort}
&&
\hspace{-10mm}
S^{(2)}_{\rm HDG} = \frac{V_3}{2}  \sum_{k, k^\prime \geq 1}^{\infty}
\biggl[
\left(M(k,k^\prime) + M(k,-k^\prime)+ M(-k,k^\prime)+ M(-k,-k^\prime) \right) a_k a_{k^\prime}
\notag\\
&&
\hspace{-10mm}
+ i\left(M(k,k^\prime) - M(k,-k^\prime)+ M(-k,k^\prime)- M(-k,-k^\prime) \right) a_k b_{k^\prime}
\notag\\
&&
\hspace{-10mm}
+ i \left(M(k,k^\prime) + M(k,-k^\prime)- M(-k,k^\prime)- M(-k,-k^\prime) \right) b_k a_{k^\prime}
\notag\\
&&
\hspace{-10mm}
+ \left(-M(k,k^\prime) + M(k,-k^\prime) - M(-k,k^\prime)+ M(-k,-k^\prime) \right) b_k b_{k^\prime}
\biggr] \, ,
\notag \\
&&
\hspace{-10mm}
= \frac{V_3}{2}  \sum_{k, k^\prime \geq 1}^{\infty}
\biggl[ 
M_{11}(k,k^\prime) a_k a_{k^\prime}
+ M_{12} (k,k^\prime) a_k b_{k^\prime}
+ M_{21} (k,k^\prime) b_k a_{k^\prime}
+ M_{22} (k,k^\prime) b_k b_{k^\prime} 
\biggr] \, .
\eea
In the path-integral given in eq. (\ref{eq:hdgpath_exp}) one has to take 
expectation value of $S^{(2)}_{\rm HDG}$. In this then the mixed 
terms appearing in $S^{(2)}_{\rm HDG}$ don't contribute as the 
action appearing in exponent is quadratic in $a_k$ and $b_k$. 
So, only the part $M_{11}(k,k^\prime)$ and $M_{22}(k,k^\prime)$
contributes. Moreover, even for these two only the terms 
for which $k=k^\prime$ contributes, others will vanish. 
These observations simplify the perturbative computations drastically. 
For $k=k^\prime$ the expressions for $M_{11}(k,k)$ and $M_{22}(k,k)$
is given by,
\begin{align}
\label{eq:M11}
M_{11}(k,k) &= \int_0^1 {\rm d}t
\biggl[
\left\{A_1 (2\pi k)^2 + A_2 (2\pi k)^4 + A_3\right\} 2 \cos (4\pi k t)
\notag \\
&
+2 \left\{-A_1 (2\pi k)^2 + A_2 (2 \pi k)^4 + A_3 \right\} \biggr] \, ,
\\
M_{22}(k,k) & = \int_0^1 {\rm d}t 
\biggl[
- \left\{A_1 (2\pi k)^2 + A_2 (2\pi k)^4 + A_3\right\} 2 \cos (4\pi k t)
\notag \\
&
+ 2 \left\{-A_1 (2\pi k)^2 + A_2 (2 \pi k)^4 + A_3 \right\} \biggr]\, .
\end{align}
At this point now we only have to perform the integrations over 
$a_k$ and $b_k$ as dictated by the path-integral in eq. (\ref{eq:hdgpath_exp}). 
This path-integral consist of two parts: the leading part is the 
Einstein-Hilbert piece which has been computed before while the 
second term is the correction term coming from the higher-derivative. 
We will compute this piece. Performing the integrations over 
$a_k$ and $b_k$, and making use of eq. (\ref{eq:Neh_exp}) we get
the following
\bea
\label{eq:akbkint}
&&
{\cal N}_{\rm EH} \int_{-\infty}^{\infty} \prod_{k=1}^{\infty} {\rm d} a_k {\rm d} b_k \left(\frac{2}{i}\right)
\biggl[
\al {\cal N}_1+ \frac{i}{\hbar} S^{(2)}_{\rm HDG} 
\biggr] \exp\left(\frac{i S^{(2)}_{\rm EH}}{\hbar}\right) 
\notag \\
&&
= \left(
\frac{3 i V_3}{4 \pi N \hbar}
\right)^{1/2} \biggl[
\al {\cal N}_1 - \frac{N}{3} \sum_{k=1}^\infty
\left\{
M_{11}(k,k) + M_{22}(k,k)
\right\} (2 \pi k)^{-2}
\biggr] \, ,
\notag \\
&&
= \left(
\frac{3 i V_3}{4 \pi N \hbar}
\right)^{1/2} \biggl[
\al {\cal N}_1 - \frac{4 N}{3} \sum_{k=1}^\infty
\int_0^1 {\rm d}t 
\left\{
A_1 + (2 \pi k)^2 A_2 + A_3 (2 \pi k)^{-2}
\right\} 
\biggr] \, ,
\notag \\
&&
= - \frac{N}{18}
\left(\frac{3 i V_3}{4 \pi N \hbar}\right)^{1/2} \int_0^1 {\rm d}t A_3 \, ,
\eea
where we have absorbed the infinite piece by defining 
the infinite constant ${\cal N}_1$ as
\beq
\label{eq:N1fixed}
{\cal N}_1 = \frac{4 N}{3\al} \sum_{k=1}^\infty
\int_0^1 {\rm d}t 
\left\{
A_1 + (2 \pi k)^2 A_2\right\} \, ,
\eeq
and $A_3(t)$ is given in eq. (\ref{eq:A3}). Putting together 
all terms we find the value of the $Q$-integration to be
\beq
\label{eq:Qint_end}
\int_{Q[0]=0}^{Q[1]=0} {\cal D} Q(t) \exp\left(\frac{i S^{(2)}_{\rm grav}}{\hbar}\right)
= \left(\frac{3 i V_3}{4 \pi N \hbar}\right)^{1/2} 
\biggl[
1 - \frac{N}{18} \int_0^1 {\rm d}t A_3 + {\cal O}(\al^2)
\biggr] \, .
\eeq
The $t$-integration here over $A_3$ is complicated and lengthy but 
can be performed using {\it Mathematica}. It carries $N$ dependence 
which is crucial in the $N$-integration.
Using that we obtain that the expression for the transition 
probability from $q[0]=a$ to $q[1]=b$.
This is given by
\beq
\label{eq:Gab_exp}
G[a,b]
= \int_{0^+}^{\infty} {\rm d} N \exp\left( \frac{iS^{(0)}_{\rm grav}}{\hbar}\right)
\left(\frac{3 i V_3}{4 \pi N \hbar}\right)^{1/2} 
\biggl[
1 - \frac{N}{18} \int_0^1 {\rm d}t A_3
+ {\cal O}(\al^2)
\biggr] \, ,
\eeq
where the form of $A_3(t)$ is given in eq. (\ref{eq:A3})
and. $S^{(0)}_{\rm grav}$ is given in eq. (\ref{eq:onshellTint}). 
After performing the $t$-integration one obtains $A_3$ 
\bea
\label{eq:IntA3t}
&&
\hspace{-5mm}
\int_0^1 {\rm d}t A_3 = \frac{\al}{9N^3} \biggl[
\frac{314928 k^2 (a-b) (a^2 + 6ab +b^2) N^2}{ab  L(N)^2}
- \frac{944784 k^2 (a-b)^3(a+b)}{ab \Lam L(N)^2}
\notag \\
&&
\hspace{-5mm}
+ \frac{5832 \left\{3(a-b)^2(a^2+b^2) - (a+b)^3 N^2 \Lam \right\}}{a^2 b^2 \Lam L(N)}
+ \frac{1458 (a^2 - b^2) \left\{72 k^2 + (a-b)^2 (3k - 2\Lam) \right\}}{ab \Lam L(N)}
\notag \\
&&
\hspace{-5mm}
+ \frac{486 (a-b) \left\{ 72 k^2 - (a^2 + 6ab +b^2)  (3k - 2\Lam) \right\} }{ab L(N)} 
- \frac{9(a-b)^2 (a+b)(9k + 26 \Lam) N^2}{a^2 b^2} 
\notag \\
&&
\hspace{-5mm}
- \frac{1944 k^2 (a^2 + b^2)}{a^2 b^2 \Lam} 
- \frac{9(9k +26 \Lam) \left\{(a^2 + b^2) (3(b-b)^2 - 2 (a+b) N^2 \Lam\right\}}{\Lam a^2 b^2} 
\notag \\
&&
\hspace{-5mm}
- \frac{54 (a-b) (3k - 2\Lam) \left\{ 3(a+b) + N^2 \Lam \right\}}{ab \Lam} 
+ \frac{3(9k +26 \Lam) (a^2 + b^2) L(N)}{\Lam a^2 b^2} 
\notag \\
&&
\hspace{-5mm}
-\frac{5(3(a-b)^2(a^2 + ab + b^2) - (a^3+b^3) N^2 \Lam) L(N)}{a^3 b^3} 
+\frac{419906 N^6 k^2 \Lam^2 +648 N^6 \Lam^2 (2\Lam -3k)}{\sqrt{-L(N)^5}}
\notag \\
&&
\hspace{-5mm}
\times 
\biggl(
\tan^{-1} \frac{3(a-b) + N^2 \Lam}{\sqrt{-L(N)}}
+ \tan^{-1} \frac{3(b-a) + N^2 \Lam}{\sqrt{-L(N)}}
\biggr)
\biggr] \, .
\eea
The integrand has singularities at $N=0$ and branch cut 
along the real-$N$ axis. This integration can be performed 
by making use of Picard-Lefschetz theory. To achieve this
we first compute the saddle-points of the above keeping 
in mind that we are doing the study perturbatively in $\al$.

\subsection{Saddles}
\label{saddles}

To compute the expression in eq. (\ref{eq:Gab_exp}) one has 
to compute all the saddles of the action $S^{(0)}_{\rm grav}$,
then apply the Picard-Lefschetz (PL) theory to include the 
contribution of the saddles to the path-integral. 
The saddles points of the path-integral can be worked out 
by looking at extrema of the action $S^{(0)}_{\rm grav}$, by varying it 
with respect to $N$. Here due to complexity of the problem 
we solve for saddles perturbatively in $\al$. The action 
$S^{(0)}_{\rm grav}$ is given in eq. (\ref{eq:onshellTint}). 
This can be varied with respect to $N$,
and plugging $N\to (\bar{N} + \al N_1)$, we obtain the following 
equations for $\bar{N}$ and $N_1$ to be
\bea
&&
\label{eq:Nbeq}
\frac{3(a-b)^2}{2 \bar{N}^2} -\Lam(a+b)+6 k+\frac{\Lam^2 \bar{N}^2}{6}=0 \, ,
\\
&&
\label{eq:N1eq}
N_1 = f(\bar{N}) \, ,
\eea
where $f(\bar{N})$ is a given in terms of $\bar{N}$ which is 
determined from the eq. (\ref{eq:Nbeq}). It is given by
\bea
\label{eq:fN1}
&&
f(\bar{N})= 
\frac{1}{4 \left(\Lam^2 \bar{N}^4-9 (a-b)^2\right)} 
\biggl[
\frac{81 (a-b)^4}{ab \bar{N}} 
- \frac{27 (a-b)^2 (a+b) \Lam \bar{N}}{ab} - 144 k \Lam \bar{N}^3 
\notag \\
&&
-\frac{3 \Lam^2 \bar{N}^3 \left(3 a^2 +3 b^2 + 8a^2b + 8ab^2 -6 ab\right)}{a b}
+ \frac{3a + 3b + 8ab}{ab} \Lam^3 \bar{N}^5
\notag \\
&&
-\frac{36 (6k+\Lam) \left\{
3(a-b)^2 + 6 k \bar{N}^2 - (a+b) \Lam \bar{N}^2 
\right\}}{\sqrt{k} }
\notag \\
&&
\times \biggl(\tan ^{-1}\frac{3 (a- b)+\Lam \bar{N}^2}{6 \bar{N} \sqrt{k}}
+\tan ^{-1}\frac{3(b-a)+ \Lam \bar{N}^2}{6\bar{N} \sqrt{k}}\biggr)
\biggr] \, .
\eea
One can solve the first equation to find $\bar{N}$. Eq. (\ref{eq:Nbeq}) is quadratic 
in $\bar{N}^2$, and hence will have four roots. Corresponding 
to each root there will be correction coming from higher-derivative
gravity which is given by $N_1$. The four roots for $\bar{N}$
are 
\beq
\label{eq:Nbroots}
\bar{N}_{\pm \pm}
=\pm \frac{\sqrt{3}}{\Lam} 
\left[
\sqrt{a\Lam -3k} \pm \sqrt{b\Lam -3k}
\right] \, .
\eeq
Corresponding to each root there is $N_1$ computed from 
$f(\bar{N})$ given in eq. (\ref{eq:fN1}). The full saddle
including the correction from higher-derivative gravity is given by
\beq
\label{eq:fullSad}
N_{\pm\pm}^{(s)} = \bar{N}_{\pm\pm} + \al F\left(\bar{N}_{\pm\pm} \right)
\eeq
It should be specified here 
that $f(\bar{N})$ doesn't have a $a\to0$ limit. In the sense 
that if we do a small $a$ expansion then the saddle has a
$1/a$ pole which is proportional to higher-derivative coupling. 
The small $a$ expansion of the saddles is 
\beq
\label{eq:smallaNs}
N_{\pm\pm}^{(s)} = \pm \frac{9\al \sqrt{-k}}{2a}
+ \left[
\pm \frac{3\sqrt{-k} \pm \sqrt{-9k + 3 b \Lam}}{\Lam}
+ {\cal O}(\al) \right] + {\cal O}(a) \, .
\eeq
The appearance of $1/a$ pole in the saddle could be artefact 
of the perturbation theory and may disappear if the computation 
is done non-perturbatively. Its appearance signals that the limit 
$a\to0$ is not trustworthy in the perturbative style of computation.
It also indicates the realm of validity of the perturbative analysis 
where $a$ shouldn't be taken to be too small or zero. As a
result in this paper to be consistent with the perturbative treatment 
we will study the scenario of $a\neq0$ only. 

Once the location of saddles are known for the given boundary conditions,
these can be plugged in the action in eq. (\ref{eq:onshellTint}) to obtain 
the corresponding on-shell action of the theory. In the present case of $R^2$ gravity
this is very lengthy expression even when one computes it perturbatively 
to first order in ${\cal O}(\al)$. This on-shell approximation is required 
for computing the saddle-point approximation. In next section we will use 
it to perform the $N$-integration using the PL-theory to obtain the transition 
probability.

\section{$N$-integration via Picard-Lefschetz}
\label{Ninnt}

Performing the $N$-integration requires making use of Picard-Lefschetz theory
and finding paths of steepest descent and ascent. Instead of performing the 
Euclideanization of path-integral by doing Wick-rotation and contour 
deformation, here in PL-theory one instead complexifies the field variable 
and perform the path-integral along the contours of steepest descent 
(known as Lefschetz thimbles) and summing over the contributions 
of all such thimbles. This powerful methodology offers a natural 
exponential damping along each thimble instead of an oscillatory integral. 

The problem of performing path-integration gets translated to a task 
of computing thimbles on a complex functional space by making use 
of Picard-Lefschetz theory. One can start with path-integral in the following manner 
\beq
\label{eq:pathmock}
I = \int {\cal D}z(t) \, e^{i {\cal S}(z)/\hbar} \, ,
\eeq
where the exponent is functional of $z(t)$. Generically this integral is 
complicated to perform as the integrand can be quite oscillatory. 
In usual quantum field theory the standard methodology is to 
Wick-rotate the above to imaginary time thereby making it 
exponentially damped. In PL theory one continue both 
$z(t)$ and ${\cal S}(z)$ in to complex plane, where one 
interprets ${\cal S}$ as an holomorphic functional of
$z(t)$ satisfying a functional form of Cauchy-Riemann 
conditions
\begin{align}
\label{eq:CRfunc}
\frac{\de {\cal S}}{\de \bar{z}} = 0 
\Rightarrow
\begin{cases}
\frac{\de {\rm Re} {\cal S}}{\de x}
&= \frac{ \de {\rm Im} {\cal S}}{\de y} \, , \\
\frac{\de {\rm Re} {\cal S}}{\de y}
&= - \frac{ \de {\rm Im} {\cal S}}{\de x} \, .
\end{cases}
\end{align}
%

\subsection{Flow equations}
\label{floweq}

If one write the complex exponential as
${\cal I} = i {\cal S}/\hbar = h + iH$ and write 
$z(t) = x_1(t) + i x_2(t)$ then downward flow 
is defined as
\beq
\label{eq:downFlowDef}
\frac{{\rm d} x_i}{{\rm d} \lam}
= - g_{ij} \frac{\pt h}{\pt x_j} \, ,
\eeq
where $g_{ij}$ is a metric defined on the complex manifold
and $\lam$ is flow parameter. The plus sign in front of 
$g_{ij}$ will correspond to steepest ascent contours which can be 
denoted as ${\cal K}_\sg$ (where $\sg$ refers to the saddle point to which it is attached), 
while negative one gives the steepest descent contour also knowns as \textit{thimbles}
and denoted by ${\cal J}_\sg$.
For the steepest descent contour the real part $h$ (also called 
Morse function) decreases monotonically as one moves 
away from the critical point along the flows. This can be seen 
by computing 
\beq
\label{eq:flowMonoDec_h}
\frac{{\rm d} h}{{\rm d} \lam}
= g_{ij} \frac{{\rm d} x^i}{{\rm d} \lam} \frac{\pt h}{\pt x_j} 
= - \left(\frac{{\rm d} x_i}{{\rm d}\lam}\frac{{\rm d} x^i}{{\rm d}\lam}\right)
\leq 0 \, .
\eeq
This holds generically for any Riemannian metric. However, for the purpose 
of the paper and simplicity we assume $g_{z,z}=g_{\bar{z},\bar{z}}=0$
and $g_{z,\bar{z}}= g_{\bar{z},z}=1/2$. This leads to simplified 
flow equations 
\beq
\label{eq:simpflow}
\frac{{\rm d}z}{{\rm d} \lam} = \pm \frac{\pt \bar{\cal I}}{\pt \bar{z}} \, ,
\hspace{5mm}
\frac{{\rm d}\bar{z}}{{\rm d} \lam} = \pm \frac{\pt {\cal I}}{\pt z} \, .
\eeq
These flow equations immediately imply that the imaginary 
part of ${\rm Im} {\cal I}=H$ is constant along the flow lines. 
\beq
\label{eq:consHflow}
\frac{{\rm d}H}{{\rm d} \lam} 
= \frac{1}{2i} \frac{{\rm d} ({\cal I} - {\cal \bar{I}})}{{\rm d} \lam} 
= \frac{1}{2i} \left(
\frac{\pt {\cal I}}{\pt z} \frac{{\rm d} z}{{\rm d} \lam} 
- \frac{\pt \bar{\cal I}}{\pt \bar{z}}\frac{{\rm d}\bar{z}}{{\rm d} \lam}
\right) = 0 \, .
\eeq
This is a boon to treatment of the oscillatory path-integral 
which gets transformed in to a convergent well-behaved 
computable quantity along the flow lines which are the 
Picard-Lefschetz thimbles. This also means that 
$H$ is conserved along the flow lines while the real 
part $h$ is a monotonically decreasing along the downward 
flow starting from critical point. 

In the complex $N$-plane the flow equations corresponding to 
steepest descent (ascent) becomes the following 
\begin{subequations}
\begin{align}
\label{eq:STdes}
& {\rm Descent} \Rightarrow 
& \frac{{\rm d} x_1}{{\rm d} \lam} = - \frac{\pt {\rm Re}{\cal I}}{\pt x_1} \, ,
\hspace{5mm}
&
\frac{{\rm d} x_2}{{\rm d} \lam} = - \frac{\pt {\rm Re}{\cal I}}{\pt x_2} \, , 
\\
\label{eq:STaes}
& {\rm Ascent} \Rightarrow 
& \frac{{\rm d} x_1}{{\rm d} \lam} = \frac{\pt {\rm Re}{\cal I}}{\pt x_1} \, ,
\hspace{5mm}
&
\frac{{\rm d} x_2}{{\rm d} \lam} = \frac{\pt {\rm Re}{\cal I}}{\pt x_2} \, ,
\end{align}
\end{subequations}
as the ${\rm Im} {\cal I}=0$ along the flow lines. 
These equations can be used to determine the trajectories of the 
steepest descent and ascent in the complex $N$-plane emanating
from the saddle point. Each saddle point has a steepest descent trajectory 
starting from it and a steepest ascent trajectory ending in it. Depending 
on boundary conditions, value of $k$, $\Lam$ and $\al$, the saddle 
points move to different locations and accordingly the trajectories 
change their shape.These equations are however generically complicated 
coupled ODEs whose solutions require making use of numerical recipes. 
The flow lines can also be determined by making use of the knowledge that 
the phase $H$ is constant along them: $H(N) = H(N_s)$. This however, 
generically gives more contour lines than the ones which are usual steepest 
descent and ascent. These extra lines are disconnected from the saddle 
points. 

Once the trajectories are known, it is then easy to find the \textit{relevant} saddle 
points by observing the flow of corresponding steepest ascent ${\cal K}_\sg$ trajectory 
intersecting the original integration contour \cite{Feldbrugge:2017kzv,Feldbrugge:2017fcc,Feldbrugge:2017mbc}.
Then the original integration $(0^+,\infty)$ becomes a summation over 
contribution from all the steepest descent contours passing through relevant saddles. 
Formally it can expressed as
\beq
\label{eq:sadsum}
(0^+,\infty) = \sum_\sg n_\sg {\cal J}_\sg \, ,
\eeq
where $n_\sg$ takes values $\pm1, 0$ depending on the relevance of 
saddles, while ${\cal J}_\sg$ here refers to integration performed 
along the steepest descent path. 
Here we follow those footsteps to analyse the case of 
$R^2$ gravity.

\subsection{Flow directions}
\label{flowDir}

The direction of flow lines emanating from the saddles or going into it can be 
determined analytically (to some extent) by expanding the action of theory 
given in eq. (\ref{eq:onshellTint}) around the saddles given in eq. (\ref{eq:fullSad}). 
If we write $N = N_s + \de N$, then the action has a power series expansion in $\de N$.
\beq
\label{eq:NexpSad}
S^{(0)} = S^{(0)}_s + \left. \frac{{\rm d}S^{(0)}}{{\rm d}N} \right|_{N=N_s} \de N 
+ \frac{1}{2} \left. \frac{{\rm d}^2 S^{(0)}}{{\rm d}N^2} \right|_{N=N_s} \left(\de N\right)^2 
+ \cdots
\eeq
The first order terms will vanish identically by definition of saddles. 
The second order terms can be obtained directly from the action 
in eq. (\ref{eq:onshellTint}) by taking double derivative with respect to $N$.
From this the direction of flows can be determined. It should be recall that 
the imaginary part of exponential $iS$ (or $H$) is constant along the flow lines.
This then implies that ${\rm Im} \left[iS - i S (N_s) \right]=0$. At the saddle point
one can write ${\rm d}^2 S^{(0)}/{\rm d}N^2 = re^{i \rho}$, as it can be a complex 
quantity depending on the boundary conditions and $\rho$ can 
be determined numerically for the given boundary conditions. Near the saddle
point the change in $H$ will go like $\D(H) \propto i 
\left(\left. {\rm d}^2 S^{(0)}/{\rm d}N^2 \right|_{N_s}\right) \left(\de N\right)^2
\sim n^2 e^{i\left(\pi/2 + 2\ta + \rho \right)}$, where 
we write $\de N = n e^{i \ta}$ and $\ta$ is the direction of flow lines. 
Given that the imaginary part $H$ remains constant along the 
flow lines, so this means 
\beq
\label{eq:flowang}
\ta = \frac{(2k-1)\pi}{4} - \frac{\rho}{2} \, ,
\eeq
where $k \in \mathbb{Z}$. For the steepest descent/ascent flow lines 
have angles $\ta_{\rm des/aes}$ when the phase for $\D H$ is such 
that correspond to $e^{i\left(\pi/2 + 2\ta + \rho \right)} = \mp1$. This implies
\beq
\label{eq:TaDesAes}
\ta_{\rm des} = k \pi + \frac{\pi}{4} - \frac{\rho}{2} \, ,
\hspace{5mm}
\ta_{\rm aes} = k \pi - \frac{\pi}{4} - \frac{\rho}{2} \, .
\eeq
These angles can be computed numerically for the given boundary conditions
and for gravitational actions. In this paper as we are approaching the problem 
perturbatively so we cannot explore all possibilities for the boundary conditions,
namely the case of no-boundary Universe ($a=0$) we will not consider. 
In the case when boundary conditions lead to real saddles, the saddle point 
action is also real, which then implies that for such boundary conditions 
$\rho=0$. We will consider this case in paper, which occurs when $(3k/\Lam)<a<b$.

\subsection{Saddle-point approximation}
\label{sadPtApp}

Once the saddle points, flow directions and steepest descent/ascent paths 
(denoted by ${\cal J}_\sg/{\cal K}_\sg$ respectively)
are figured out, it is then easy to find the relevant saddle points. 
A saddle point is relevant when the steepest ascent path emanating from it 
coincides with the original contour of integration, which in this case is
$(0^+, \infty)$. The original $N$-integration reduces to summation of
all contour integrations done along the Lefschetz thimbles. 

Then we make use of saddle point approximation to compute the 
transition amplitude by solving the eq. (\ref{eq:Gab_exp}). In the 
$\hbar\to0$ limit we have
\bea
\label{eq:Gabapprox}
&&
G[a,b] = \sum_\sg n_\sg 
\left(\frac{3 i V_3}{4 \pi \hbar}\right)^{1/2} \int_{{\cal J}_\sg} \frac{{\rm d}N}{\sqrt{N}}
\exp\left( \frac{iS^{(0)}_{\rm grav}}{\hbar}\right)
\biggl[
1 - \frac{N}{18} \int_0^1 {\rm d}t A_3
+ {\cal O}(\al^2)
\biggr] \, ,
\notag \\
&&
\approx
\sum_\sg n_\sg 
\left(\frac{3 i V_3}{4 \pi \hbar}\right)^{1/2}
\exp\left( \frac{iS^{(0)}_{\rm Saddle}}{\hbar}\right)
\frac{1}{\sqrt{N_s}}
\biggl[
1 - \frac{N_s}{18} \int_0^1 {\rm d}t A_3 (N_s)
+ {\cal O}(\al^2) 
\biggr] 
\notag \\
&&
\times \int_{{\cal J}_\sg}
{\rm d}N \exp \left(
\frac{i S^{(0)}_{NN}}{\hbar} (N - N_s)^2
\right)\biggl(1 + {\cal O}(\sqrt{\hbar}) \biggr) \, ,
\eea
where we consider only leading term in the $\hbar$ expansion. Here 
$S^{(0)}_{\rm Saddle}$ is the on-shell action computed at the saddle point,
and $S^{(0)}_{NN}$ is the second variation of the action at the saddle point $N_s$
which is given in eq. (\ref{eq:fullSad}). 
The action at the saddle point is given by
\bea
\label{ea:S0sadact}
&&
\hspace{-10mm}
S^{(0)}_{\rm Saddle}
= 6 \sqrt{k} c_1 \biggl[
\frac{2k}{\Lam} \left(U^3 - c_2 V^3 \right)
+ 2 k \al \left\{U(9 + U^2) - c_2 V(9+V^2) \right\}
+ \Lam \al \biggl\{\frac{3U + 2 U^3}{1+U^2} 
\notag\\
&&
\hspace{-10mm}
- \frac{(3V + 2V^3) c_2}{1+V^2} \biggr\}
+ 3\al (2k + \Lam) \left( \tan^{-1} U - c_2 \tan^{-1} V \right)
+ {\cal O}(\al^2)
\biggr] \, ,
\eea
where $c_{1,2}={+,-}$, and we have defined variables 
\beq
\label{eq:UVvar}
a = \frac{3k(1+U^2)}{\Lam} \, , 
\hspace{5mm}
b = \frac{3k(1+V^2)}{\Lam} \, .
\eeq
In terms of variables $U$ and $V$ the saddle points acquire a compact form and 
are given by
\bea
\label{eq:SadPtUV}
&&
N_s = - \frac{3 c_1 \sqrt{k} (U - c_2 V)}{\Lam} 
- \frac{3 c_1}{2 \sqrt{k}} \biggl[
(U - c_2 V) \biggl\{
k \left(2 + \frac{4c_2}{UV} \right)
+ \frac{(1 - c_2 UV) \Lam}{(1+U^2)(1+V^2)} 
\biggr\}
\notag \\
&&
+ (6k +\Lam) \left( \tan^{-1} U - c_2 \tan^{-1} V \right)
+ {\cal O}(\al^2)
\biggr] \, .
\eea
The second variation of the action at the saddle point is given by,
\bea
\label{eq:S0NNsadpt}
&&
\left. S^{(0)}_{NN} \right|_{N_s}= \frac{4 c_1 \sqrt{k} U V \Lam}{U - c_2 V}
+ \frac{2 c_2 \Lam^2 \al}{\sqrt{k} (U - c_2 V)}
\biggl[
\frac{4k(U^2 V^2 - U^2 - V^2 - c_2 UV)}{UV}
\notag\\
&&
+ \frac{U^2V^2 (-c_2 + UV)}{(1+U^2) (1+V^2)}
\biggr]
- \frac{2 \Lam^2 \al}{\sqrt{k} (U - c_2 V)^2} \biggl[
2k(6 c_2 UV + 6U^2 + 6 V^2 +5 U^2 V^2) 
\notag \\
&&
+ \Lam (2 c_2 UV + 2U^2 + 2V^2 +U^2 V^2) 
\biggr]  \left( \tan^{-1} U - c_2 \tan^{-1} V \right)
+ {\cal O}(\al^2) \, .
\eea
If we write $N-N_s=n e^{i \ta}$, where $\ta$ is the angle the Lefschetz thimble make 
with the real $N$-axis, then the above integration can be performed easily. 
It gives the following 
\bea
\label{eq:GabTrans}
G[a,b] \approx 
\sum_\sg n_\sg 
\sqrt{\frac{3 i V_3}{4 N_s \lvert S^{(0)}_{NN} \rvert}}
\exp\left(i \ta_\sg + \frac{iS^{(0)}_{\rm Saddle}}{\hbar}\right)
\biggl[
1 - \frac{N_s}{18} \int_0^1 {\rm d}t A_3 (N_s)
+ {\cal O}(\al^2) 
\biggr] \, .
\eea
This is a generic expression and is true for various boundary conditions 
in the saddle point approximation to first order in higher-derivative coupling. 
The higher-derivative corrections are contained in the exponential 
and the $A_3$-term in the square bracket. This is the new correction 
term in the transition amplitude which arises due to the higher-derivative 
coupling, which in this case is $R^2$ modified gravity. Similar 
kind of corrections can be computed perturbatively for other type of 
UV modification of gravity. In the next sub-section we will compute the 
above amplitude for a particular case of boundary condition.

\subsection{$(3k/\Lam)<a<b$}
\label{cuni}

This boundary condition correspond to classical case. The 
saddle points lie on real axis in complex $N$ plane. It is easy to 
see from eq. (\ref{eq:Nbroots}) that the roots $\bar{N}_{\pm\pm}$
are real for the case when $(3k/\Lam)<a<b$. In terms of $U$ and $V$ this 
boundary condition means that $0<U<V$. In this case 
$f(\bar{N})$ following from eq. (\ref{eq:fN1}) is also 
real, thereby resulting in a set of real saddle points. 
For each of this real saddle point the action corresponding to them is 
real. Moreover, the second variation of the action at the saddle
point is also real thereby resulting in $\rho=0$. 
This has interesting implication for the directions of 
steepest ascent and descent. 
\beq
\label{eq:StDeAe}
\ta_{\rm des} = k \pi + \frac{\pi}{4} \, ,
\hspace{5mm}
\ta_{\rm aes} = k \pi - \frac{\pi}{4} \, .
\eeq
This is nice simplification that is achieved in the case of the 
boundary condition $(3k/\Lam)<a<b$. We numerically compute the 
flow-lines by solving the coupled differential equations given in 
eq. (\ref{eq:STdes} \& \ref{eq:STaes}) for various values of $\al$. 
The flow lines can also be computed by making use of the condition 
$H(N)=H(N_s)$. However, this process gives additional flow lines: 
lines which are disconnected from the saddle points. 

For this kind of boundary conditions there are four real saddle points:
two positive and two negative ones. 
The two positive ones lie on the original contour of integration $(0^+, \infty)$. 
This is similar to the situation in Einstein-Hilbert case where the saddle 
point also lie on real $N$-axis for this set of boundary conditions
\cite{Feldbrugge:2017kzv}. The two saddle points lying on the positive 
real $N$-axis ($0<N_s^{(1)}<N_s^{(2)}$) are the relevant saddle points. 
The steepest descent paths ending in them 
is used to construct Lefschetz thimbles. This immediately gives the 
transition amplitude $G[a,b]$ in the saddle point approximation 
by making use of eq. (\ref{eq:GabTrans}). The various ingredients 
entering the expression can be computed by making use of eq. 
(\ref{ea:S0sadact}, \ref{eq:SadPtUV}, \ref{eq:S0NNsadpt} and \ref{eq:IntA3t}),
while the phase $\ta_{\rm des}^{(1)} = -\pi/4$ and $\ta_{\rm des}^{(2)} = \pi/4$. 
It is then straightforward task to perform the summation in eq. (\ref{eq:GabTrans}) 
giving a lengthy expression for $G[a,b]$. Due to its length we avoid writing it here.

\section{Conclusion}
\label{conc}

In this paper we investigate Lorentzian quantum cosmology in the 
context when gravity gets modified in the deep ultraviolet (UV). 
It is a complicated task to define the path-integral of quantum gravity 
while taking care of issues regarding gauge-fixing, renormalizabilty, 
regularization, functional measure and most importantly finding a 
contour of integration along which it is absolutely convergent. 
Once all this ingredients are in place, one can proceed further 
to make an attempt in computing meaningful quantities like 
transition amplitude. Although a mathematically challenging task to ask is the probability 
of transition from one geometry to another, in the mini-superspace 
approximation this problem gets simplified a bit as the problem reduces 
to a one-dimensional path-integral. This simplification is justified 
as at very large distance (cosmological scales) our Universe is homogenous and isotropic.
Such maximally symmetric geometry also occurs at very early stages of our 
Universe. In this simplified picture one can address meaningful question 
regarding the situation of spacetime geometry at very early Universe 
and transition probability from one $3$-geometry to another. 
This probability has been computed in the context of Einstein-Hilbert gravity
in euclidean path-integral and the outcome was no-boundary proposal 
of Universe  \cite{Hawking:1981gb,Hartle:1983ai,Hawking:1983hj}, 
and tunnelling proposal \cite{Vilenkin:1982de,Vilenkin:1983xq,Vilenkin:1984wp}.
Recently effort has been made to tackle the gravitational path-integral 
directly in Lorentzian signature by making use of Picard-Lefschetz theory
\cite{Feldbrugge:2017kzv,Feldbrugge:2017fcc,Feldbrugge:2017mbc},
which allows to choose a contour of integration uniquely.

In this paper we address the above problem in the context of 
$R^2$-modified gravity, where our motivation comes from the fact that 
gravity is expected to get UV modified and Einstein gravity is 
not the full picture in UV. Criterion of renormalizabilty in four spacetime 
dimensions further motivates one to treat EH gravity as an effective field 
theory which follows from some UV complete more fundamental theory.
Within the field theory framework requirements of renormalizability 
hints at modifying Einstein-Hilbert gravity in UV. This 
usually is done by modifying the field propagator in UV, 
which naturally happens in fourth order higher-derivative gravity
(adding $R^2$ and $R_\mn R^\mn$ to Einstein-Hilbert gravity)
\cite{Stelle:1976gc,Narain:2011gs,Narain:2012nf}
and/or by considering non-localities 
\cite{Modesto:2011kw,Modesto:2017sdr}.
This pushes us to ask the question how such modifications will 
shape our understanding of quantum cosmology? We make 
baby steps in understanding this and a first attempt 
in computing transition probability from one $3$-geometry 
to another in the mini-superspace approximation 
in the $R^2$-gravity model.

We follow the strategy described in \cite{Halliwell:1988ik,Halliwell:1989dy,Feldbrugge:2017kzv}
to analyse the path-integral of $R^2$-gravity in the mini-superspace 
approximation. The action for this in gauge $\dot{N}=0$ is given 
in eq. (\ref{eq:GravActMinisup}). We do the analysis in Jordon frame
directly. Although in section \ref{hdgms} we do mention about transforming 
the $R^2$-model to Einstein frame, but this strategy is specific to only 
$F(R)$ models of gravity and cannot be extended to other type of 
modified gravity, for example $R_\mn R^\mn$ modification. In Jordon 
frame we analyse the equation of motion which is seen to be highly 
non-linear. We approach the problem perturbatively and find the solution to 
equation of motion to first order in $\al$, which allow us to find the saddle points 
of the system to first order in $\al$. As the solution to equation of motion doesn't 
satisfy the constraint (\ref{eq:constNq}), so one can write the generic solution 
$q(t) = q_b (t) + Q(t)$. Accordingly one gets a path-integral over $Q(t)$.
This we perform perturbatively to first order in $\al$, as unlike in the 
Einstein-Hilbert gravity it cannot be performed exactly. We finally write 
an expression for transition probability to ${\cal O}(\al)$ in eq. (\ref{eq:Gab_exp}) 
which now involves only an integration over lapse $N$. 

We analyse the integration over lapse $N$ by making use of Picard-Lefschetz 
theory. The perturbative approximations allows us to make progress in this direction.
In the saddle point approximation we perform the $N$-integration along the 
appropriately chosen contour dictated by Lefschetz thimbles. 
In the saddle point approximation we obtain an expression for the
transition amplitude from one $3$-geometry to another in the $R^2$-gravity to ${\cal O}(\al)$, 
which is mentioned in eq. (\ref{eq:GabTrans}). This is the main result of the paper.

It is valid in the perturbative approximation when $\al$ is small.
We apply this in a simple example when the boundary 
conditions satisfy $(3k/\Lam)<a<b$. This is the classical case
where the saddle point lie in real axis. The qualitative picture of 
location of saddles and steepest descent/ascent lines in the complex-$N$
plane remains the same as in EH gravity case. In the small-$\al$ 
approximation we are unable to study the no-boundary proposal
as the approximation is unreliable in early Universe regime. It will be nice 
if the whole computation of transition amplitude can be done without 
resorting to small-$\al$ approximation. We leave that to future publication.

\bigskip
\centerline{\bf Acknowledgements} 

We will like to thank Jean-Luc Lehners for useful discussions. 
GN will like to thank Nirmalya Kajuri and Avinash Raju for discussion. 
GN is supported by ``Zhuoyue" (distinguished) Fellowship (ZYBH2018-03).
H. Q. Z. is supported by the National Natural Science Foundation of China (Grants
No. 11675140, No. 11705005, and No. 11875095).

\appendix


\end{document}